\documentclass[11pt]{article}
\usepackage{graphicx}
\usepackage[margin=1.25in]{geometry}
\usepackage[usenames,dvipsnames]{color}
\usepackage{url}
\usepackage[colorlinks = true,
            linkcolor = blue,
            urlcolor  = blue,
            citecolor = blue,
            anchorcolor = blue]{hyperref}
\usepackage{subcaption}

\newcommand\pubdate{\today}

\def\Title#1{\begin{center} {\LARGE #1 } \end{center}}
\def\Author#1{\begin{center}{ \sc #1} \end{center}}
\def\Address#1{\begin{center}{ \it #1} \end{center}}

\newcommand\pubblock{\rightline{\begin{tabular}{l} 
         \pubdate \end{tabular}}}
\newenvironment{Abstract}{\begin{quotation} \begin{center}
                       ABSTRACT
     \end{center}\bigskip  }{\end{quotation}}

\def\beq{\begin{equation}}
\def\eeq#1{\label{#1}\end{equation}}
\def\eeqn{\end{equation}}

\newenvironment{Eqnarray}%
   {\arraycolsep 0.14em\begin{eqnarray}}{\end{eqnarray}}
\def\beqa{\begin{Eqnarray}}
\def\eeqa#1{\label{#1}\end{Eqnarray}}
\def\eeqan{\end{Eqnarray}}

\let\bar=\overbar

\def\lsim{\mathrel{\raise.3ex\hbox{$<$\kern-.75em\lower1ex\hbox{$\sim$}}}}
\def\gsim{\mathrel{\raise.3ex\hbox{$>$\kern-.75em\lower1ex\hbox{$\sim$}}}}

\def\del{\partial}
\def\Dslash{\not{\hbox{\kern-4pt $D$}}}
\def\dslash{\not{\hbox{\kern-2pt $\del$}}}
\def\pslash{\not{\hbox{\kern-2pt $p$}}}
\def\ETmiss{\not{\hbox{\kern-4pt $E$}}_T}

\def\Dlr{\mathrel{\raise1.5ex\hbox{$\leftrightarrow$\kern-1em\lower1.5ex\hbox{$D$}}}}

\def\MSB{{\bar{M \kern -2pt S}}}
\def\msb{{\bar{\scriptsize M \kern -1pt S}}}

\def\drb{{\bar{\scriptsize D \kern -1pt R}}}

\newcommand\snowmass{\begin{center}\rule[-0.2in]{\hsize}{0.01in}\\\rule{\hsize}{0.01in}\\
\vskip 0.1in Submitted to the  Proceedings of the US Community Study\\ 
on the Future of Particle Physics (Snowmass 2021)\\ 
\rule{\hsize}{0.01in}\\\rule[+0.2in]{\hsize}{0.01in} \end{center}}

\begin{document}

\pubblock

\Title{Analysis Cyberinfrastructure: Challenges and Opportunities}

\bigskip 

\Author{K. Lannon, P. Brenner, M. Hildreth, K. Hurtado Anampa, A. Malta Rodrigues, K. Mohrman, D. Thain, B. Tovar}

\medskip

\Address{University of Notre Dame, Notre Dame, IN 46556, USA}

\medskip

\begin{Abstract}
\noindent Analysis cyberinfrastructure refers to the combination of software and computer hardware used to support late-stage data analysis in High Energy Physics (HEP).  For the purposes of this white paper, late-stage data analysis refers specifically to the step of transforming the most reduced common data format produced by a given experimental collaboration (for example, nanoAOD for the CMS experiment) into histograms.  In this white paper, we reflect on observations gathered from a recent experience with data analysis using a recent, python-based analysis framework, and extrapolate these experiences though the High-Luminosity LHC era as way of highlighting potential R\&D topics in analysis cyberinfrastructure.
\end{Abstract}

\snowmass

\def\thefootnote{\fnsymbol{footnote}}
\setcounter{footnote}{0}
%


\section{Executive Summary}

Based on our experiences performing an analysis of CMS data, starting from the nanoAOD data format, using data collected from 2016-2018 with the CMS detector (Run 2) as well as the corresponding simulated data, we conclude that there are no fundamental limitations that would prevent a complete analysis iteration (from nanoAOD to histograms) from being completed in roughly 10 minutes.  However, we note that this hypothetical target is not met with the current analysis infrastructure.  Reflecting on the reasons for this and considering other related observations, we construct the following list of potentially interesting R\&D topics in the area of analysis cyberinfrastructure:
\begin{itemize}
    \item Distributed storage systems.
    \item Integration between different layers in the software stack.
    \item Specialized facilitator training.
    \item Monitoring, especially across software stack layers.
    \item Fault tolerance and recovery across software stack layers.
    \item Automated tuning of workflow parameters based on information gathered during processing.
    \item Approaches for addressing resource contention between users to preserve fast turn-around time of individual analysis iterations.
\end{itemize}

\section{Introduction}

With the upcoming high luminosity run of the LHC (HL-LHC), the LHC physics community is in danger of becoming victims of its own success.  Motivated by the need to probe increasingly rare physics processes, HL-LHC will deliver luminosity to the experiments at roughly an order of magnitude higher rate than previously.  The detectors will be upgraded to collect data at higher rates as well, leading to a significantly larger volume of data for analysts to sift through.  While many aspects of LHC data processing should scale up reasonably well without major rethinking, late-stage data analysis will require more care.  This is because in the late-stages of data analysis, progress occurs iteratively, with humans strongly coupled in the loop.  The speed of performing one pass through the data quickly becomes the limiting factor to the pace of exploration and innovation.  Already, the time to make one pass through the LHC Run 2 dataset for most analyses has grown to be hours or even a day or more.  Scaling this up by a factor or ten or more would bring the pace of science to a grinding halt.  The challenge facing the HEP community is to find a way to preserve or even improve the pace at which data analysis iterations can be performed in the face of a rapidly growing volume of data.

Some discussion of terminology is in order.  In this white paper, we use the term ``analysis cyberinfrastructure'' rather than the more common ``analysis facility'' to emphasize the belief that integration of both hardware and software will be required.  While the term ``analysis'' may have a variety of meanings within the broader field of HEP and even within HEP subfields (e.g. LHC physics), here by ``analysis'' we specifically mean the the activities necessary to transform individual collision data in whatever is the most reduced form provided centrally by a given experiment into histograms and other similar statistical summaries.  To give a concrete example, for the CMS experiment, this would encompass going from the nanoAOD format to histograms.  Certainly there are many other activities that could also justifiably be construed as analysis, such as performing specialized event reconstruction, performing fits, training machine learning methods, etc., but we will not tackle those topics here.

Defined in this way, analysis may consist of up to two specific steps:
\begin{description}
\item[Skimming:]  This step refers to the selection of a specific subset of event data that can be determined to be of interest with some relatively straightforward selection criteria.  Put another way, skimming is the process of discarding all events that you can easily determine you have no interest in scrutinizing further.
\item[Histogramming:]  This steps involves potentially deriving quantities from the raw event data and then reducing that data (or some subset) into a statistical representation like a histogram or something similar.  
\end{description}
It is not necessarily the case that these two steps will be implemented as discrete activities that can be decoupled.  They may in fact be interleaved in a way that prevents running them separately.

\section{Analysis Model} \label{sec:AnalysisModel}

In order to discuss the challenges and opportunities in a concrete way, it will be helpful to have a specific model of analysis.  For this white paper, we will use an analysis model based on recent experiences of a specific analysis group based at the University of Notre Dame.  This analysis group is analyzing data from the CMS experiment starting from the nanoAOD format, which is a ``flat'' ROOT tree data format centrally maintained by the CMS experiment.  The analysis software used by the group is based on the Coffea~\cite{coffea} analysis framework scaled out using the Work~Queue~\cite{workqueue} back-end.  Coffea is written in python and relies on the underlying scientific python ecosystem, including numpy~\cite{numpy} and matplotlib~\cite{matplotlib} as well as some more specialized python packages for HEP, notably awkward~\cite{awkward} and uproot~\cite{uproot}.  It is worth noting that Coffea follows a numpy-like approach with an emphasis on performing calculations on ``columns'' of data loaded from the nanoAOD ROOT files with the columns held in memory.  While we think it's fair to say that this is an emerging trend in HEP data analysis, it also engenders complications that will impact some of our conclusions in this white paper.  Analysis models based on other programming styles, like the more traditional ``nested-loops'' construction may lead to different conclusions than what we present here.

For this white paper we will focus primarily on insights gleaned from following this single analysis group, although at the end of the document, we will also discuss some considerations for scaling up to support multiple analysis groups.  An important parameter for this discussion is the amount of data that the group is analyzing.  Table~\ref{tab:dataVolume} shows the amount of data used by this analysis\footnote{For the curious, this analysis involves selecting a sample of top quark collisions in a final state involving multiple charged leptons and subjecting that sample to a data analysis similar to the one described in~\cite{TOP-19-001}}.   The analysis underway uses data collected by the CMS experiment during LHC's Run 2 (2016--2018), but Table~\ref{tab:dataVolume} also includes estimates of total data volume that might be expected by the end of LHC Run 3 (2025) and the HL-LHC ($\sim 2040$).  These estimates are obtained simply by scaling the data volume up by a factor of two and twenty respectively, based on on rough projections of the amount of data expected to be recorded by the CMS detector during those periods.  Clearly, there are many complicating factors that are being glossed over here that might reduce (software improvements, more effective selection strategies, etc.) or increase (greater complexity of data coming from additional overlapping collisions, known as pile-up, expected for later running) the projected amounts of data relative to these estimates.  The numbers here are intended simply to set the order order of magnitude of the data expected, and should not be taken as precise predictions.

The data tallied in Table~\ref{tab:dataVolume} includes both the actual data recorded by the CMS detector as well as the appropriate simulated data needed to perform the data analysis.  The ``unskimmed'' column represents the amount of data in the full nanoAOD datasets with no preselection or reduction applied.  The ``skimmed'' column shows the data remaining after applying some fairly loose selection criteria and only saving the events passing.  Typically, the analysts can work with the skimmed data volume as the requirements were inclusive enough that all desired events were selected in the skim.

\begin{table}
\begin{center}
    \begin{tabular}{l|r|r}
    \hline\hline
         Period & Unskimmed & Skimmed \\
         \hline
         Run 2 & 11.5\,TB & 2.5\,TB \\
         Run 3 & 23\,TB & 5\,TB \\
         HL-LHC & 230\,TB & 50\,TB \\
    \hline\hline
    \end{tabular}
\end{center}
\caption{The data used in the reference analysis considered in this white paper (labeled ``Run 2''), as well as estimates for the data volume for future periods obtained by scaling by factors of two (for Run 3) and twenty (for HL-LHC).}
\label{tab:dataVolume}
\end{table}

The other salient factor required to model analysis activities is some measure of the CPU processing resources required.  To make it easy to combine with the data volumes provided above, we estimated that our analysis code can process 1\,GB of data in approximately 3.3 minutes using a single core.  This estimate was obtained using Intel Xeon CPUs E5-2680 v3 @ 2.50GHz with 128GB of RAM.  This analysis software performs a variety of tasks, including selecting events and objects (electrons, muons, jets) within these events that pass certain selection criteria, considering combinations of objects (electron or muon pairs that comprise a Z boson and jet triplets representing a hadronic top quark decay), and calculating derived quantities connected with the individual objects or combinations, and filling histograms with those quantities.   While certainly this just represents one possible example of analysis code, we argue that it is reasonably representative, falling in between the extremes of a very simple analysis (e.g. one that simply analyzes collisions containing lepton pairs from Drell-Yan production) and very complicated and computationally intensive analyses (e.g. one that computes matrix element probabilities or evaluates a computationally expensive deep neural network for each event).  Nevertheless, one should keep in mind that depending on where an analysis falls along this spectrum, the conclusions presented here may vary.

\section{Analysis Turn-Around Time: The 10-Minute Challenge}

The scientific process is driven by iterative exploration, cycling between making hypothesis and testing them, trial and error.  In the context of ``analysis'' as defined here, one iteration represents making a change to the analysis, as expressed in code, and then using the updated code to process the data into histograms.  The amount of time that it takes to complete one iteration therefore represents a key limiting factor on the pace of scientific innovation.  On the other hand, obtaining the shortest possible analysis turn-around time requires analyzers to invest effort into optimizing their analysis workflow, which is effort not spent on physics analysis.  It has been our observation that the equilibrium between these two considerations has, over a broad range of time, resulted in analysis iteration times in the range of a few hours to ``overnight.''  Is it possible to do significantly better than this?

Let's start with an admittedly arbitrary target of 10 minutes to complete one iteration.  That is to go from the reduced common format (i.e. nanoAOD) to histograms.  We will consider the question both for the skimmed and the unskimmed scenarios outlined above, since the skimming requirements change infrequently enough to allow the skimmed sample to constitute a stable starting point for analysis.  The first consideration we will examine involves the I/O rates needed to attain this 10-minute goal.  Those rates are summarized in Table~\ref{tab:dataRates}.  The table includes the necessary rates not just for a 10-minute turn-around time, but also for longer times.  Clearly, even for the skimmed Run 2 dataset, a 10-minute turn-around time is beyond the capabilities of standard hard-drive performance.  However, using a distributed storage cluster should provide the necessary aggregated bandwidth.  For the Run 2 case, assuming each server in the storage cluster can perform at 5 Gb/s\footnote{Assumes a multi-disk storage server provisioned with standard consumer grade HDs.}, seven servers should be able to supply the necessary bandwidth for the skimmed case, twice that many servers could provide a 20-minute turn-around time for the unskimmed case.  At the HL-LHC scale, using performance numbers for today's hardware, a 10-minute turn-around time might not be feasible.  However, for skimmed data, it would be possible to obtain a 1-2 hour turn-around time with a reasonably sized (11-22 server) storage cluster.  Depending on future pricing, using higher performance storage options, such as SSD or NVMe storage coupled with higher internet speeds should allow even faster turn-around times.  So, while the challenge is far from trivial, the goal of 10-minute turn-around time seems achievable from an I/O rate standpoint.

\begin{table}
\begin{center}
    \begin{tabular}{|r|rrr|rrr|}
        \hline\hline
        & \multicolumn{3}{|c|}{Unskimmed} & \multicolumn{3}{|c|}{Skimmed} \\
        Iteration Time & Run 2 & Run 3 & HL-LHC & Run 2 & Run 3 & HL-LHC \\
         \hline
        10\,min.  & 150\,Gb/s & 310\,Gb/s & 3,100\,Gb/s & 33\,Gb/s & 65\,Gb/s & 650\,Gb/s \\
        30\,min.  &  51\,Gb/s & 100\,Gb/s & 1,000\,Gb/s & 11\,Gb/s & 22\,Gb/s & 220\,Gb/s \\
        60\,min.  &  26\,Gb/s &  51\,Gb/s &   510\,Gb/s &  5\,Gb/s & 11\,Gb/s & 110\,Gb/s \\ 
        120\,min. &  13\,Gb/s &  26\,Gb/s &   260\,Gb/s &  3\,Gb/s &  5\,Gb/s &  54\,Gb/s \\
    \hline\hline
    \end{tabular}
\end{center}
\caption{Aggregate I/O rates for the storage system storing the analysis data required to achieve the given turn-around time.  The data volumes correspond to the relevant entries in Table~\ref{tab:dataVolume}.}
\label{tab:dataRates}
\end{table}

The next aspect to consider in hitting this 10-minute goal is CPU concurrency.  Using the figure of 3.3 minutes to process 1\,GB of data explained in Section~\ref{sec:AnalysisModel}, we can calculate the amount of concurrency required to process the data from Table~\ref{tab:dataVolume} in 10 minutes.  The results of this calculation are in Table~\ref{tab:cpuConcurrency}.  For the Run 2 dataset, the numbers shown in this table for processing skims are well within reason for a large Tier-3 site and would represent only a fraction of the size of a Tier-2 site.  Even for the HL-LHC, the required CPU resources shouldn't be prohibitive.  It should be noted that for one analysis iteration, these resources would only be used for 10 minutes, making this an ideal use-case for leveraging opportunistic resources, including processing resources claimed by job pilots that have unclaimed slots but that are too close to the end of their lifetime to match to more conventional processing jobs.  The I/O needs of a given analysis processing task should be reasonably modest, on the order of 40\,Mb/s meaning that a single compute host can easily support multiple processing tasks.  In fact, the more challenging requirement, based on experiences with the analysis group at Notre Dame and the processes software, as described in Section~\ref{sec:AnalysisModel}, is likely the memory requirement.  Individual processing tasks use anywhere from 1\,GB to 5\,GB. The upper-end of this range tends to result from a rather memory-intensive custom histogram implementation required by that specific analysis.  It should be noted that the processing in a Coffea task, to date, is always single-threaded.  If Coffea were to support multithreaded processing, which should be reasonably straightforward to implement through the numpy-like data structures on which Coffea is based, then the memory requirement could be significantly reduced, allowing multiple processing threads to share a common memory allocation for the memory-heavy histograms.

\begin{table}
\begin{center}
    \begin{tabular}{|r|rrr|rrr|}
        \hline\hline
        & \multicolumn{3}{|c|}{Unskimmed} & \multicolumn{3}{|c|}{Skimmed} \\
        Iteration Time & Run 2 & Run 3 & HL-LHC & Run 2 & Run 3 & HL-LHC \\
         \hline
        10\,min.  & 3,800 & 7,600 & 76,000 & 810 & 1,600 & 16,000 \\
        30\,min.  & 1,300 & 2,500 & 25,000 & 270 &   540 &  5,400 \\
        60\,min.  &   630 & 1,300 & 13,000 & 130 &   270 &  2,700 \\ 
        120\,min. &   320 &   630 &  6,300 &  70 &   130 &  1,300 \\
    \hline\hline
    \end{tabular}
\end{center}
\caption{The number of CPUs that must be used in parallel to process the data listed in Table~\ref{tab:dataVolume} in the stated iteration time.}
\label{tab:cpuConcurrency}
\end{table}

Given that the possibility of a 10-minute iteration time seems achievable with the current Run 2 datasets, it is interesting to ask how close we've come to hitting that mark at Notre Dame.  The Notre Dame Tier-3 provides roughly 1,000 CPU cores of dedicated processing resources, and has opportunistic access to another 30,000 CPU cores hosted at Notre Dame's Center for Research Computing.  At any given  moment, up to 30\% of the CRC resources are idle and available for opportunistic use.  All computing resources are provisioned with at least 1 Gb/s ethernet connections and most have between 4--5 GB/core of memory.  Notre Dame's storage cluster consists of fifteen servers hosting a combined storage of roughly 800\,TB of storage, with each server equipped with 10\,Gb/s ethernet.  The storage cluster uses the HDFS filesystem and is accessible over the LAN via XRootD.  The XRootD infrastructure consists of a redirector node and data servers running on five of the fifteen storage server nodes.  Notre Dame also deploys on XCache infrastructure consisting of a redirector and five caching proxies, each of which has 24-72 TB cache storage and a 10\,Gb/s ethernet connection to the WAN.

All told, the above resources should be sufficient to achieve a 10-minute turn-around time (or close to it) per analysis iteration.  However, the best actual performance observed so far has been approximately 2-3 hours for one analysis iteration.  The performance gap is not fully understood at the time of this writing, but it seems to be connected to performance of the HDFS or XRootD software layers of the storage infrastructure, possibly related to the metadata handling or overhead processing network transactions.  Arriving at this level of performance required diagnosing and improving bottlenecks located in several other elements of the software stack before hitting this bandwidth limitation in the storage software layer.  The experience of working to achieve optimal performance factors into the observations made in subsequent sections of this white paper.

\section{Complexity of the Software Stack}

\begin{figure}
\begin{subfigure}{.4\textwidth}
  \centering
  \includegraphics[width=.8\linewidth]{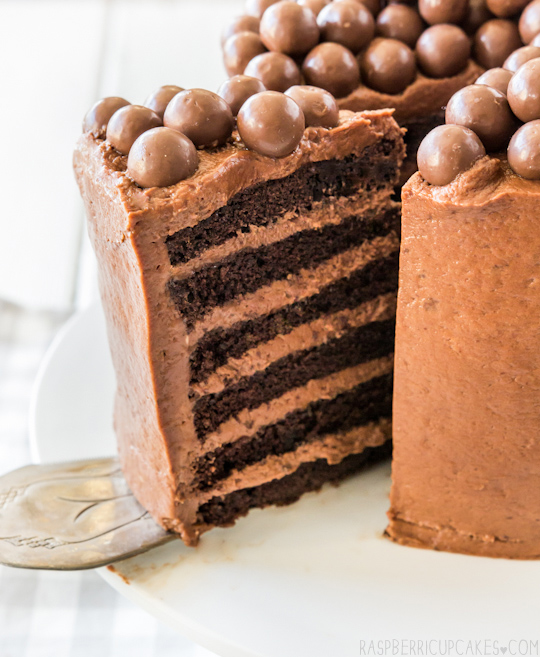}
  \caption{Photo courtesy of~\cite{raspberricupcakes}}.
  \label{fig:NiceCake}
\end{subfigure}%
\begin{subfigure}{.6\textwidth}
  \centering
  \includegraphics[width=.8\linewidth]{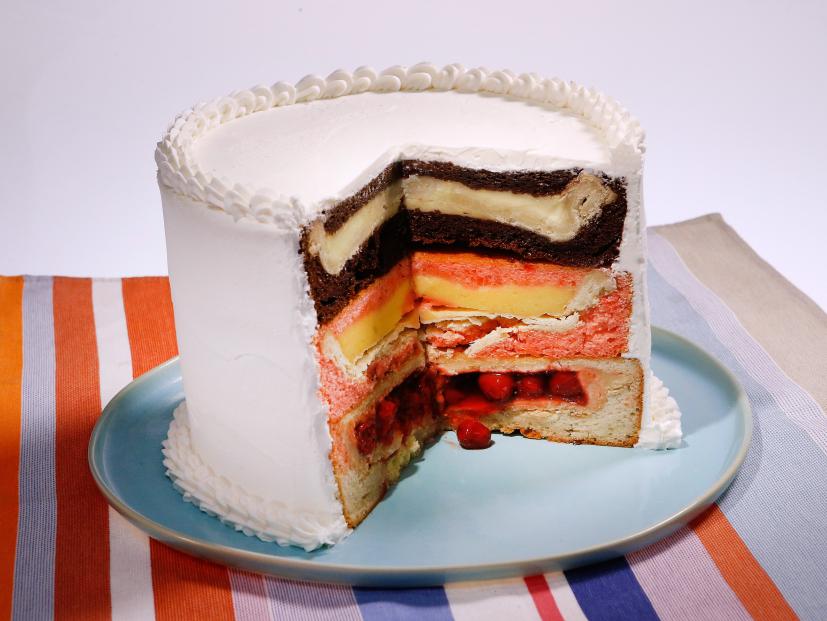}
  \caption{Photo courtesy of~\cite{FoodNetwork}}
  \label{fig:PieCaken}
\end{subfigure}
\caption{The analysis facility software stack can be seen as a ``layer cake,'' but is it more accurate to view it as a chocolate layer cake as in~(\subref{fig:NiceCake}) or a ``PieCaken'' as in~(\subref{fig:PieCaken})?}
\label{fig:Cakes}
\end{figure}

Cyber technologists like to borrow common words and concepts like ``architecture'' and ``cloud'' to try to make complex systems more understandable to society. The term ``layer cake'' has been used similarly in everything from cybersecurity to operating systems to computational workflows.  While this nomenclature is helpful, it is also potentially dangerous as the human experience is more apt to see the layer cake in example Figure~\ref{fig:NiceCake} than the more cyber accurate layer cake in example Figure~\ref{fig:PieCaken}. While the layer cake in Figure~\ref{fig:NiceCake} gives both the chef and the diner a well defined process and experience, the layer cake in Figure~\ref{fig:PieCaken} is fraught with culinary challenges for both the chef and the diner.  In the HEP analysis community, we have users with a very nuanced palette that requires an analysis facility much more like the layer cake in Figure~\ref{fig:PieCaken} (known as a ``PieCaken''); we must therefore set the user expectations and engineering design priorities accordingly.  Finally, while scaling up additional layers or consumers of the Figure~\ref{fig:NiceCake} cake is straight forward; scaling up the PieCaken with additional layers will undoubtedly have unexpected taste outcomes and scaling up the user base and workload types will undoubtedly result in users and workloads that simply do not align with the very unique characteristics of the facility.

\begin{figure}
\begin{center}
\includegraphics[width=0.9\textwidth]{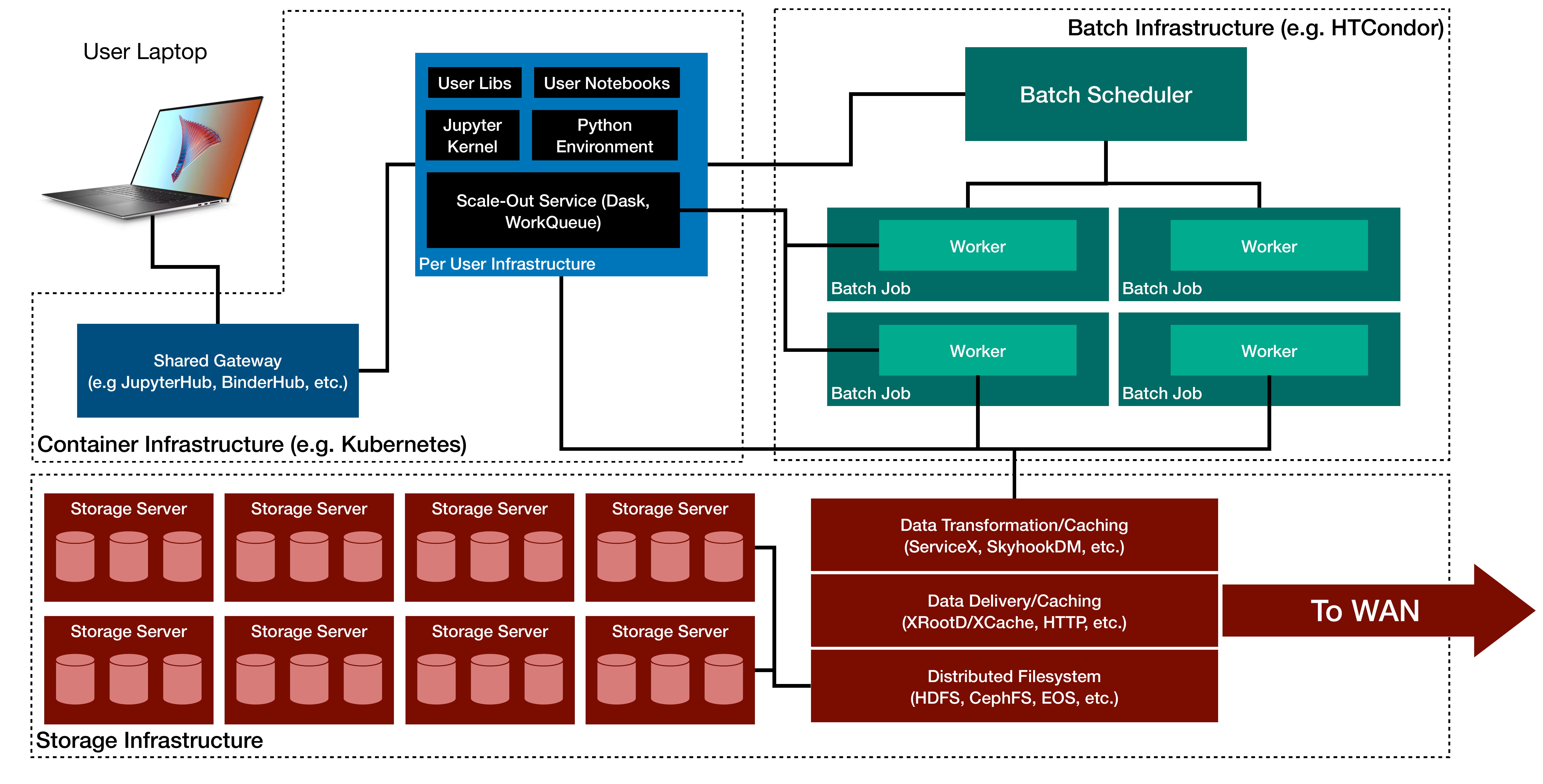}
\end{center}
\caption{A schematic of an analysis.  This diagram represents the overall vision intended for the Notre Dame analysis cyberinfrastructure, although currently not all aspects are implemented.  These details are included in the text.}
\label{fig:AFDiagram}
\end{figure}

A more technical diagram representing the ``layer cake'' of typical HEP analysis cyberinfrastructure is shown in Figure~\ref{fig:AFDiagram}.  This diagram makes it clear that there are numerous layers that must act in concert to achieve the desired performance of an analysis iteration.  The layers can be roughly classified as follows:
\begin{description}
\item[Analysis Software:] This includes the software the the analysts write themselves along with any direct dependencies.  User analysis software should not be viewed as simple.  It may take the form of one or more user created libraries (installed via cloning repositories or pip) as well as one more more applications built on top of those libraries, and also libraries from the broader scientific python ecosystem, including potentially machine learning libraries.  At Notre Dame, this includes the TopCoffea software~\cite{TopCoffea} which consists of both a library component as well as applications.
\item[Analysis Framework:] This layer sits directly underneath the analysis software to facilitate the interaction between the analysis software and the common reduced data format (e.g. nanoAOD) as well as providing some connection to scale-out mechanisms. At Notre Dame, this includes Coffea~\cite{coffea} which provides an interface to the nanoAOD data through awkward~\cite{awkward} and uproot~\cite{uproot}.  The interface to Work~Queue~\cite{workqueue} is also used as a mechanism for scaling out the analysis workflow at Notre Dame.
\item[User Interface:]  This layer is responsible for providing users with the ability to interact with the computational resources.  At Notre Dame, this layer consists of an interactive linux login node that also provides a Jupyterhub instance.  In the future this might be enhanced by or replaced with a more extensive Jupyterhub or Binderhub infrastructure managed via Kubernetes across multiple servers.
\item[Batch Infrastructure:] This layer connects the distributed application and framework to the computational resources to perform the analysis.  It includes both the batch system itself as well as any software that interfaces to the batch system to generate jobs.  It should be noted that in some cases, the submit node to the batch system may not be the same node as the interactive node for users on the cluster.  At Notre Dame, the batch system is HTCondor (two pools, one dedicated and one opportunistic each accessible via a different submit node) and a Work~Queue factory.
\item[Storage Infrastructure:]  Cyberinfrastructure pertaining to storage consists of a stack with potentially several layers.  The top-most layer may consist of applications that transform data from one format to another, potentially also caching the results of that transformation.  Examples of such software include ServiceX~\cite{servicex} and SkyhookDM~\cite{skyhookdm}.  Below that one finds a data delivery layer, such as XRootD or HTTP which provides a link between higher layers in the system and the distributed lower layers.  The data delivery system may be accessed directly by the distributed analysis application or via the transformation layer described above.  The data delivery layer may also provide access to data over the WAN as well as caching services.  Below the delivery layer sits the distributed file system layer.  This layer presents a unified view of the distributed storage as if it were a single file system.  Options include HDFS, CephFS, or EOS.  At Notre Dame, our storage infrastructure currently has no transformation layer.  We use XRootD for data delivery through two redirectors: one for our local storage and one for our XCache system.  Our distributed file system layer is currently HDFS, but we are investigating changing to CephFS or EOS for performance reasons.  The HDFS storage cluster consists of fifteen data servers totaling almost 800\,TB of raw storage while the XCache system consists of five caching proxy servers providing a total of about 300\,TB of cache space.
\end{description}

While this multi-layered software approach has numerous benefits, there are challenges that should not be ignored.  Especially when it comes to obtaining optimal performance, bottlenecks can occur at any layer in the stack.  Because the stack encompasses the full range from user written analysis code to relatively low-level software associated with the storage and compute infrastructures, it is unlikely that any one person associated with the cyberinfrastructure will have a sufficiently broad expertise to identify and resolve any possible bottlenecks.  Changes at one layer can manifest unintended consequences at other layers.  

Although these challenges should not be ignored or underestimated, there are a number of opportunities for mitigating them.  From the cyberinfrastructure perspective, a focus on integration across the layers can provide important advantages when trying to diagnose potential bottlenecks.  To this end, monitoring should not be omitted from the design of the analysis infrastructure.  While monitoring is often approached layer-wise, with the low-level cyberinfrastructure monitoring delegated to systems administrators while higher levels primarily serve users, a vertically integrated monitoring approach that aggregates information from across the layers provides a better approach and should be attainable with industry-standard tools already compatible with many of the layers our software stack (e.g. Prometheus, Loki, Logstash, Elasticsearch, Grafana, and Kibana).  Another important opportunity for addressing these challenges comes in human form, through training.  While training of users and administrators involved with the analysis cyberinfrastructure is very important, we should also not neglect the possibility of training specific facilitators, who understand the full stack from the user level to the systems administrator level to serve as liaisons and troubleshooters.  Such trained experts could find a home either within specific experiments (e.g. CMS, ATLAS, etc.) or in broader organizations such as WLCG or OSG, to provide assistance in diagnosing and tuning the software stack of multiple analysis facilities.  Combining integrated monitoring with specially trained analysis facilities experts would provide a significant advantage in the quest to obtain optimal performance.

Another aspect which spans the software stack and impacts the ability to achieve the desired end-to-end throughput is the resilience of each interconnected layer.  Failures, especially transient ones, are an unavoidable fact of life in distributed computing.  However, it is often difficult for a given software layer to decide how to respond to a failure.  For sporadic failures, retries, especially with appropriate delays or fallback mechanisms, are essential for achieving reliable performance at scale.  However, retries and fallbacks in one layer can also mask persistent problems in another layer while slowing overall throughput to a crawl.  To achieve reliably high performance while also being able to respond to and fix problems as they arise, the individual layers in the infrastructure need to be able to communicate about error mechanisms, and any information about encountered errors needs to be able to percolate up to a level at which humans (users, sysadmins, or ideally both) have visibility.

\section{Managing Variability in Analysis Tasks}

Another significant challenge facing analysis cyberinfrastructure is contending with the potentially broad range of analysis workflows.  While we have focused on one particular set of analysis workflows that we argue can be taken as representative, we should not ignore that there are plenty of possible variations on that theme.  For instance, the chosen example workflow is based on the nanoAOD data format from CMS, but not every analysis can start from that format.  How do things change if an analysis requires a less reduced data format?  The representative analysis has as a goal the production of histograms for statistical analysis, but what about workflows that output instead a dataset of individual events, perhaps for use in machine learning training?  Even our representative example can display significant variations in resource requirements and performance needs.  One analysis processor used as part of the representative analysis activity focuses on extracting and caching global information, like the sum of all MC weights for each simulated sample under various systematics scenarios.  This processor does much less computation than the representative processor while also reading a smaller amount of data from each processed file.  On the other hand, by flipping a few switches, the standard analysis processor can be put in a mode which performs substantially more computation and furthermore is extremely memory intensive.  It is easy to imagine (and we have in fact observed) that an optimal configuration of analysis cyberinfrastructure for one of the possibilities outlined above will likely not be optimal for others.  Experience also suggests that its not feasible for users to do extensive optimization for each possible variation in their analysis activities to ensure they always make optimal choices.  Therefore, without some strategy to address these variations, obtaining the desired performance for actual analysis deployed within any specific analysis infrastructure seems unlikely.

Fortunately, again, there are a number of strategies than can be pursued to meet this challenge.  One often overlooked option is to design analysis cyberinfrastructure with an intentionally limited scope.  The more variety of workflows supported by analysis cyberinfrastructure, the more problematic it is to provide optimal performance across the whole range of workflows.  Allowing the designers of analysis cyberinfrastructure to focus on a narrower range of use cases, opens the possibility of using strategies that wouldn't apply across the broader spectrum of possibilities.  Of course, this suggests that it may be necessary to create multiple analysis cyberinfrastructure software stacks, each targeting a particular subset of use cases.  While this might seem like a losing proposition, the fact that the underlying building blocks are highly modular and therefore likely to be useful in more than one cyberinfrastructure stack, plus the increased possibility of using targeted automatic optimization in the more narrowly defined use-case sets may ultimately outweigh the costs of developing multiple special-purpose cyberinfrastructure stacks.

Within the specific analysis use-case of going from most-reduced event data format (e.g. nanoAOD) to histograms, we have observed a number of possible automatic optimisation strategies that seems promising.  For example, within the Coffea framework, one parameter that has a significant impact on performance is the ``chunk size'' used in processing the data.  Chunk size is defined as the maximum number of events that the analysis processor will attempt to analyze in a single task.  (Smaller amounts of data may be used in practice because the Coffea framework will not combine date from multiple files into a single chunk.)  There are multiple competing factors that must be accounted for when tuning chunk size.  Chunk size impacts the memory requirements of the processing as larger chunk sizes imply more data will be held in memory simultaneously.  Large chunk sizes also limit the maximum amount of processing concurrency as individual chunks are processed in a single-thread.  Choosing a smaller chunk size can avoid the issues previously mentioned, but choosing a chunk size that is too small will lead to other problems.  A small chunk size partitions the dataset into many tasks, multiplying any per task overhead.  Having a large number of chunks puts more load on the components tasked with tracking the processing tasks and generates more work in aggregating the results of the processing.  Leaving the choice of chunk size to users often results in poor analysis throughput and sometimes even in analysis workflows that cannot be completed.  However, we have found that it's possible to implement an algorithm to dynamically scale the chunk size during an analysis run to meet a pre-defined run-time or memory usage target or a combination of the two~\cite{topeft-ipdps}.  With this auto-scaling approach, chunk size can be adjusted based on the observed statistics of already completed tasks, and if the chunk size becomes so large that it exhausts the resources provisioned for processing, the failed tasks can be automatically retried with smaller chunk size.  While this algorithm cannot match the performance obtained using an optimally-tuned chunk size estimated a priori, the loss of performance relative to that benchmark may be an acceptable trade-off for the performance boost obtained for the average user who is freed from the need to exhaustively benchmark and tune prior to each analysis run.

\section{Multi-User Considerations}

So far, this white paper has focused on considerations relevant for cyberinfrastructure used by a small number of individual users, largely because those are the situations that we have directly explored at Notre Dame.  Nevertheless, it will be important to consider how these observations and strategies evolve as we scale the analysis cyberinfrastructure up to support many users operating in an uncoordinated fashion.  There are two main aspects that we consider as we scale up the cyberinfrastructure: data volume and user activities.

For data volume, at first glance, the situation is not so challenging.  The total data volume used by the reference analysis studied in this white paper represents roughly 10\% of the total nanoAOD data volume for CMS Run 2.  Assuming that ratio holds and scaling up by a factor of twenty for HL-LHC, this means the ultimate data volume for HL-LHC to store all of the common reduced format data for a given experiment would be on the order of 2.5\,PB.  It would certainly be feasible to host the entire unskimmed common reduced dataset at a single site, and even having multiple distributed copies should be possible.  A bigger challenge will be to provide resources for skims.  If there is no coordination of skimming activities, skims will scale not just by the data volume, but also by the number of separate analysis groups supported by the analysis facility.  Assuming the size of the skims for the reference analysis considered here are representative, then if more than five groups independently create skims from a given subset of the data, the total skim volume will exceed the unskimmed volume.  Of course, there should be a large overlap of data selected by different skims, and accounting for that overlap has the potential to reduce the amount of storage that would need to be allocated.  Perhaps the most promising avenue to explore would be to replace the concept of manually generated skims with a more automated process of caching specific subsets of data as they're accessed.  

Scaling for the processing resources is more complicated to estimate.  If we focus for the moment on the Run 2 scale of analysis, a facility with 1,000 CPU cores should enable one research group to turn an analysis iteration around in 10 minutes.  Then, naively, assuming an 8 hour work day, this hypothetical facility could support 50 complete analysis iterations per day.  While this is certainly true in terms of raw capacity, it misses the subtleties that arise if there are multiple analysis groups operating in an uncoordinated fashion.  In other words, if we were to assume that the 50 iterations represented 10 different analysis groups performing 5 complete passes per day, there is no reason to believe that those passes would align nicely in time to ensure the groups would never be contending for resources.  As soon as two group try to access the same resources at the same time, the 10-minute turn-around time will be compromised as the system resources have to be split, delivering, for example, a 20 minute turnaround time to two groups working in parallel.  While, on some level, this might be unavoidable, it would be good to explore strategies for mitigating this effect.  However, these strategies will need to be coordinated across the full ``layer cake'' of the cyberinfrastructure, rather than using the conventional approach of independently addressing resource contention at the computing and storage layers (e.g. via a ``fair share'' metric).   One possible way to achieve this sort of cross-layer coordination would be to employ a central manager process that receives and evaluates processing requests holistically, prioritizing the completion of whole workflows quickly rather than just maximizing average throughput across all workflows.

Another open question to consider when exploring approaches to scaling up analysis cyberinfrastructure to support multiple groups is whether it makes sense to focus on the approach of scaling up through a specially-designed physical facility to support multiple analysis groups, to approach analysis cyberinfrastructure as an add-on feature to existing computational facilities (such as LHC Tier-2 sites), or to design analysis cyberinfrastructure to be deployed in a cloud-like approach across a variety of different physical resources from experiment owned sites like LHC Tier-1, 2, and even 3 facilities, to resources that are used on a more temporary basis, such as HPC facilities or even commercial cloud resources.  While it may be difficult to find a definitive answer to this question today, it is worth noting that following the best practices for providing analysis cyberinfrastructure as a service (i.e. using containerization and container orchestation to automate the deployment and operation) will leave open the possibility of deploying across non-conventional resources while still providing benefit even to deployments at custom-desitned and purpose-built facilities.  

\section{Conclusion and Outlook}

It is no doubt clear that without investments in new approaches to late-stage data analysis, on the HL-LHC time scale there is a danger of scientific progress stagnating as the time to complete one analysis iteration grows with the data volume past the limits of human attention spans and patience.  However, hopefully this white paper has successfully argued that there are a number of promising avenues along which to tackle this challenge.  Furthermore, many of these promising approaches would also lead to short-term benefits that would substantially improve analysis productivity even in the present day.


\bibliographystyle{JHEP}
\bibliography{myreferences}  

\end{document}